\documentclass[twocolumn,epsfig,pre,10pt,floatfix,amsmath,amssym,showpacs]{revtex4}
 
\usepackage{amsmath}
\usepackage{graphicx}
\usepackage{graphics,graphicx}
\usepackage{epstopdf}

\newcommand{\rv}{{\bf r}}
\newcommand{\kv}{{\bf k}}
\newcommand{\Gv}{{\bf G}}

\begin{document}

\title{Ordered Phases of Diblock Copolymers in Selective Solvent}
\author{Gregory M. Grason}
\affiliation{Department of Physics and Astronomy, University of California at Los Angeles, Los Angeles, CA 90024, USA}

\date{\today}

\begin{abstract}
We propose a mean-field model to explore the equilibrium coupling between micelle aggregation and lattice choice in neutral copolymer and selective solvent mixtures.  We find both thermotropic and lyotropic transitions from face-centered cubic to body-centered cubic ordered phases of spherical micelles, in agreement with experimental observations of these systems over a broad range of conditions.  Stability of the non-closed packed phase can be attributed to two physical mechanisms:  the large entropy of lattice phonons near crystal melting and the preference of the inter-micelle repulsions for the body-centered cubic structure  when the lattice becomes sufficiently dense at higher solution concentrations.  Both mechanisms are controlled by the decrease of micelle aggregation and subsequent increase of lattice density as solvent selectivity is reduced.  These results shed new light on the relationship between micelle structure -- ``crewcut" or ``hairy" -- and long-range order in micelle solutions.
\end{abstract}

\maketitle

\section{Introduction}

Well known for their ability to self-assemble into a seemingly endless array of nanoscale structures \cite{discher_eisenberg_science_02, jain_bates_science_03}, neutral block copolymer and solvent mixtures have recently received significant attention as a test bed for phenomena of a more classical nature:  the crystallization of spherical particles \cite{mcconnell_gast_prl_93, mcconnell_gast_pre_96, bang_lodge_prl_02, lodge_macro_02, bang_lodge_jcp_04, bang_lodge_prl_04_1, bang_lodge_prl_04_2, stellbrink_jphys_04, stellbrink_prl_05}.  In selective solvent copolymer chains aggregate into spherical micelles -- among other structural motifs -- protecting the insoluble polymer blocks within a core domain while allowing the soluble blocks to dissolve in a diffuse coronal domain.  At low concentrations when inter-micelle contact is rare, correlations are fluid-like.  However, at sufficiently high concentrations the swollen coronal brushes of neighboring micelles come into contact, and to minimize the repulsive effect of this contact, the micelles adopt lattice configurations.   Based on a series of experimental studies of lattice-forming copolymer micelle solutions, a basic thermodynamic picture has emerged:  generically, so-called ``crew cut" micelles characterized by short coronal brushes favor the face-centered cubic (fcc) structure, while ``hairy" micelles with larger coronae favor the body-centered cubic (bcc) lattice arrangement \cite{mcconnell_gast_prl_93, mcconnell_gast_pre_96, lodge_macro_02}.  

At low temperatures, many repulsive particle systems favor close-packed lattice structures -- such as fcc -- primarily due to the more distant nearest neighbors and subsequent weaker repulsions in these lattices.  As temperature is raised, the entropic contribution to the free energy from lattice fluctuations eventually becomes comparable to repulsive contributions.  Due to a relative abundance of soft-phonon modes of bcc lattice, this structure is particularly stable near to the crystal melting transition \cite{hoover_jcp_72}.  Indeed, quite generic consideration suggests that the bcc structure is {\it always} preferred near to the lattice melting transition due to its large entropy \cite{alexander_mctague_prl_78}.  Entropic considerations, however, are not necessary to drive a fcc-to-bcc structural transition in systems of purely repulsive particles.  When repulsions are characterized by some length scale -- as in Yukawa systems -- energetic considerations alone show that the generic preference for close-packed structures at low particle densities gives way to a preference for bcc at high densities which persists even at zero temperature \cite{hone_jcp_83,  kremer_prl_86, robbins_jcp_88}.  Indeed, the bcc lattice is well-known to be the three-dimensional ground state for Coulomb repulsions, when interactions exist over an infinite range \cite{fuchs_prs_35, carr_pr_65}.

In a series of systematic studies, Lodge and coworkers identified fcc-to-bcc transitions in copolymer micelle solutions which are both {\it thermotropic} \cite{bang_lodge_prl_02, bang_lodge_prl_04_1} -- driven by increasing temperature, $T$ --  and {\it lyotropic}  \cite{lodge_macro_02, bang_lodge_jcp_04} -- driven by increasing polymer concentration, $\Phi$.  The former transition scenario might suggest an entropic stabilization of the bcc structures due to lattice fluctuations, while the latter scenario seems to implicate a density-driven transition in which positional entropy does not play a crucial role.  In this article, we construct a mean-field model to probe the relationship between the thermodynamics of micelle aggregation and the physical mechanisms underlying structural phase transitions of micelles in copolymer solutions at fixed concentration and temperature.  While simulations of ordered structures in star-polymer solutions \cite{likos_lowen_prl_99, likos_jphys_02} performed at fixed aggregation number, $p$, and number density, $n$, shed some light onto the related problem of copolymer micelles in solutions, the thermodynamics of the present system has the complication that intra-micelle energetics must be in equilibrium with inter-micelle effects.  That is, in copolymer solutions as $\Phi$ and $T$ are varied, the system simultaneously adjusts the mean values of $n$ and $p$ as well as the structural organization of micelles.  

To address this issue, the present model combines treatments of good and bad solvent polymer domains with an effective theory for the free-energy cost -- from micelle interactions and lattice entropy -- of confining micelles within a lattice.  Therefore, we address, at a mean-field level, how the thermodynamics of the internal degrees of freedom of the constituent ``particles" of this system, the micelles, drive the changes in inter-micelle order observed on larger length scales.  The principle consequence of increasing temperature in these systems is to decrease the solvent preference for coronal polymer domain leading to a decrease in the number of chains per aggregated micelle, $p$ \cite{bang_lodge_prl_02}.  We argue below that the dissolution of micelles with increased $T$ has two effects:  to {\it reduce} the strength of the effective micelle repulsions and to {\it increase} the packing density of micelles.  Both effects tend to enhance the stability of the bcc structure as solvent selectivity is decreased, although the two mechanisms are of qualitatively different origins.  In the first case, occurring at low $\Phi$, lattice entropy stabilizes the bcc phase for sufficiently weak micelle interactions, while in the second case, at higher concentration, it is the micelle potential itself which prefers the bcc structure over fcc at high micelle densities.  Varying the relative lengths of the core and coronal polymer blocks within the model, we investigate the stability of ordered fcc and bcc phases as a function of the chain architecture.  While the ``crewcut" vs. ``hairy" micelle distinction roughly describes ordered phase behavior, we find that the relative sizes of corona and core do not wholly determine lattice choice.  Both thermotropic and lyotropic fcc-to-bcc transitions are predicted over a broad range of polymer architectures.

This article is organized as follows.  In the Sec. \ref{sec: model} we construct a mean-field model for the free-energy of the internal state of the micelles, as well as the cost of lattice confinement.  We present results for this model in Sec. \ref{sec: results} and conclude with a brief discussion in Sec. \ref{sec: discussion}. 

\section{Free Energy Model for Lattice Micelle Phases}

\label{sec: model}

In this section we construct an effective mean-field model to compute the free energy of lattice micelle phases of copolymer solutions at fixed temperature and polymer fraction, $\Phi$.  We first consider the energetics of isolated micelles in solution before turning to the inter-micelle effects.  Note that the model for isolated micelle energetics presented here has been recently analyzed in detail in the limit of low concentration (in the absence of inter-micelle effects) by Zhulina {\it et al.} \cite{zhulina_macro_05}.  We present a complete discussion here for the purpose of elucidating the results of the full calculation, which takes into account effects of lattice confinement.

\subsection{Internal Micelle Configurations}

We consider a solution of diblock copolymers of $N$ total monomers of volume fraction $\Phi$.  The lengths of the insoluble core and soluble coronal blocks are $N_c$ and $N_b$ respectively.  Temperature dependence enters the model through the selectivity of the solvent for the coronal brush domain over the core domain.  Here, we assume that solvent is excluded from the core domain entirely, so that the insoluble polymer blocks are exposed to unfavorable solvent contact only at the interface separating the core and coronal domains.  Microscopic solvent-monomer interactions give rise to a surface tension, $\gamma$, for the core-corona interface.  We generically expect the importance of molecular scale interactions to decrease with temperature, and hence, $\gamma$ to be a {\it decreasing} function of $T$ as is typically the case for example, in immiscible polymer blends \cite{bates_fredrickson_arpc_90}.  In this model the effect of solvent selectivity will be encoded entirely in $\gamma$.  The free energy of an isolated micelle configuration, $F_0$, is the sum of three parts,
\begin{equation}
F_0=F_c+F_{int}+F_b \ ,
\end{equation}
where $F_c$ and $F_b$ are the free energies of the core and coronal brush domains and $F_{int}$ is the free energy cost of exposing the core domain to solvent along the inter-domain surface.

\begin{figure}[b]
\resizebox{2.25in}{!}{\includegraphics{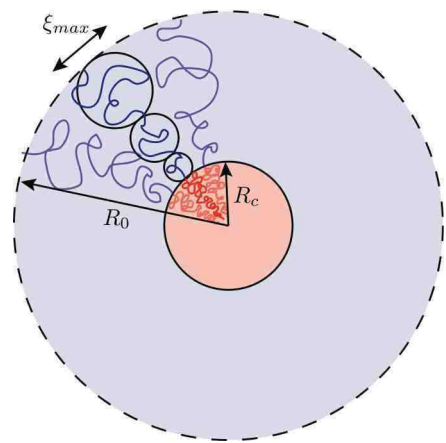}}
\caption{A cartoon of an isolated micelle in solution.  Solvent is excluded from a molten polymer core of radius, $R_c$.  A dissolve coronal brush extends from $R_c$ to a radius of $R_0$.  In the Daoud and Cotton model (see text) this brush can be divided into correlation ``blobs" of size, $\xi$, determined by the distance between neighboring chains. }
\label{fig: micelle}
\end{figure}

Consider an isolated micelle composed of $p$ total chains shown in Figure \ref{fig: micelle}.  The core domain extends to a radius, $R_c$, while the tips of the coronal brush reach to a radial distance, $R_0$.  Because solvent is absent from the core, the polymer blocks that constitute this domain are described by the statistics of the {\it molten polymer brush} \cite{degennes}.  As such, the free energy of this domain can be calculated in the strong-stretching approximation for spherical, molten polymer brushes \cite{semenov_jetp_85},
\begin{equation}
\label{eq: Fc}
\frac{F_c}{k_B T} = \frac{3 \pi^2 p}{80} \frac{R^2_c}{N_c a^2} \ , 
\end{equation}
where $a$ is the mean monomer-monomer distance.  Since excluded-volume interactions between chains are screened in the molten polymer state, the free energy of this domain is purely due to the entropic (Gaussian) cost of extending the chains in the brush.  Given the surface tension between core and coronal brush domains, we have,
\begin{equation}
\label{eq: Fint}
F_{int} = 4 \pi \gamma R_c^2 \ .
\end{equation}
Treating the core block as hydrophobic, we expect that $\gamma$ is on the order of $10 \ {\rm mN}/{\rm m}$ \cite{stellbrink_prl_05}, which in thermal units corresponds to $\gamma a^2/k_B T \approx0.5$, assuming a segment size of $a=5 {\rm \AA}$. Note that from the volume constraint on the core we have,
\begin{equation}
p=\frac{4 \pi R_c^3 \rho_0}{3N_c} \ ,
\end{equation}
where $\rho_0$ is the monomer density.

The coronal domains of the micelles are under good-solvent conditions, leading to a dramatic swelling of these blocks in comparison to the core domain.  We treat the free energy of this domain within the Daoud and Cotton model of ``star-like," spherical polymer brushes \cite{daoud_cotton_jphys_82}.  This model has been studied extensively elsewhere \cite{witten_pincus_macro_86, birshtein_zhulina_poly_89}; we present a detailed discussion here since the physics of polymer corona has important implications for the micelle-micelle interactions.  Under good-solvent conditions, the properties of a semi-dilute polymer solution are governed largely by a single length scale, $\xi$, the so-called blob size, which is determined by the mean distance between neighboring chains \cite{degennes}.  At shorter length scales than $\xi$ the correlations of the chains obey the Flory scaling, $g^3 \approx \rho_0 \xi^5 / a^2$, where $g$ is the number of monomers per blob.  For a spherical brush of $p$ chains, the average lateral distance between neighboring chains is determined by the geometrical condition that $4 \pi r^2= p \xi^2(r)$ (see Figure \ref{fig: micelle}) so that,
\begin{equation}
\xi(r)=\sqrt{\frac{4 \pi}{p}} r \ .
\end{equation}   
The local concentration of monomer in the dissolved brush is determined from $\phi_b(r)=g(r)/\xi^3(r) \propto p^{2/3}/r^{4/3}$.  Because the total number of monomers in the corona is fixed, integrating this concentration profile over the volume of the coronal brush allows us to solve for the size of the brush,
\begin{equation}
p N_b = 4 \pi \int_{R_c}^{R_0}dr~r^2 \phi_b(r) \ .
\end{equation}
Inverting this condition we find that the size of the micelle obeys,
\begin{equation}
R_0^{5/3}=R_*^{5/3}+R_c^{5/3} \ , 
\end{equation}
with
\begin{equation}
R_*^5=\frac{125}{108 \pi} \frac{ p N_b^3 a^2}{\rho_0} \ ,
\end{equation}
and $R_*$ is the size of a star polymer of $p$ chains of length $N_b$ in the limit that $R_c \rightarrow 0$.  Note that in the ``hairy" micelle case, when $R_* \gg R_c$, we have $R_0 \sim N_b^{3/5} p^{1/5}$, whereas for ``crew-cut" micelles, when $R_* \ll R_c$, we have $R_0 \sim (N_c p)^{1/3}$.

The free energy of the good solvent domain is computed by counting the total number of these thermal blobs, each of which represent a $k_B T$ worth of energy.  This is most conveniently accomplished by integrating the blob density, $\xi^{-3}(r)$, over the coronal domain volume, from which it can be shown,
\begin{equation}
\label{eq: Fb}
\frac{F_b}{k_B T} = \frac{p^{3/2}}{\sqrt{ 4 \pi }} \ln \Big(\frac{R_0}{R_c} \Big) \ . 
\end{equation}
In the absence of any inter-micelle effects, the scaling behavior of the mean aggregation number can be obtained by considering the ``hairy" and ``crew-cut" limits of $F_0$ \cite{birshtein_zhulina_poly_89}.  When $R_0\gg R_c$, the free energy of the coronal brush, $F_b$, dominates over the core stretching energy, $F_c$, so that $p \sim \gamma^{6/5} N_c^{4/5}$.   In the opposite crew-cut limit, $F_b \ll F_c$, the mean aggregation grows as $p \sim \gamma N_c$.  Decreasing solvent selectivity (decreasing $\gamma$), decreases the mean number of chains per micelle, yielding a roughly linear dependence of chain aggregation on the core-corona surface tension.

\begin{figure}
\resizebox{3in}{!}{\includegraphics{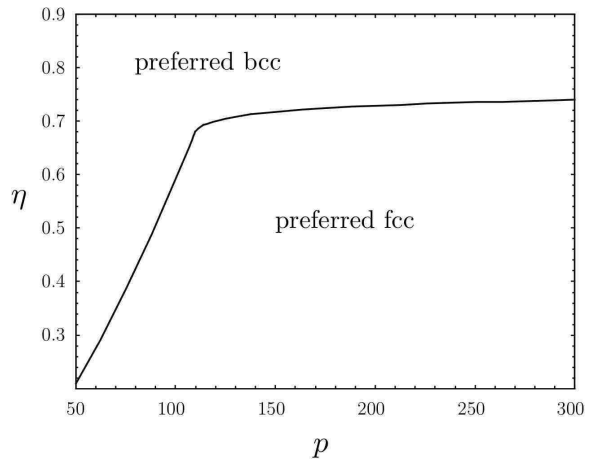}}
\caption{The solid line shows the crossover in the preference of $U(r)$ for the bcc and fcc lattice structures as function of micelle aggregation, $p$, and reduced density, $\eta=(\pi/6) n \sigma^3$.}
\label{fig: Upref}
\end{figure}

\subsection{Micelle Repulsion and Lattice Entropy}

At low polymer concentrations, the density of micelles is sufficiently small that inter-micelle effects may be ignored \cite{johner_physica_91}.  As $\Phi$ is increased well beyond the critical micelle concentration the coronal domains of neighboring micelles begin to overlap, leading to strong repulsive effects.  To minimize the free-energy cost of such repulsions, micelle solutions adopt periodic arrangements, alternately bcc or close-packed structures.  We consider a lattice of micelles of aggregation number, $p$, at number density, $n$.  Given the low solubility of single, unaggregated chains in solution \cite{johner_physica_91, dormidontova_macro_99}, we assume a negligible (zero) concentration of free chains.  Therefore, a mean micelle aggregation number of $p$ implies a micelle density,
\begin{equation}
\label{eq: n}
n=\frac{\rho_0 \Phi }{p N}\ . 
\end{equation}
Note that as micelles dissolve and $p$ decreases, the micelle density {\it increases} to maintain a fixed polymer fraction, $\Phi$.

When neighboring micelles begin to overlap at high densities (i.e. $n \gtrsim (2 R_0)^{-3}$), we must account for the repulsive effects between coronal brushes.   By a scaling analysis, Witten and Pincus showed that the pair potential for star-like, spherical brushes exhibits a logarithmic interaction at short distances \cite{witten_pincus_macro_86}.  Physically this is justified by the demand that the free-energy cost of bringing two micelles of $p$ chains each to a separation of order $2 R_c$, should be roughly the same as the free energy of creating a single micelle of $2p$ chains (see eq. (\ref{eq: Fb})).  On these grounds, Likos and coworkers proposed the following form of the pair potential for star polymers of $p$ chains \cite{likos_lowen_prl_98},
\begin{equation}
\label{eq: Ur}
\frac{U(r)}{k_B T} = \frac{ 5}{18}p^{3/2}  \left\{\begin{array}{ll} -\ln \big( \frac{r}{\sigma} \big) + \frac{1}{1+\sigma/\xi_{max}} & r \leq \sigma \\ \\
\frac{\sigma/r}{1+ \sigma/\xi_{max}}e^{ - (r-\sigma)/\xi_{max} } & r > \sigma \end{array}\right . \ .
\end{equation}
At shorter separations than a soft-core size, $\sigma$, the repulsion has the Witten and Pincus form.  Since the micelles have a finite size, at separations beyond $\sigma$, $U(r)$ has a Yukawa form, which falls off exponentially on a length scale set by the largest blob size, $\xi_{max}$ (see Fig. \ref{fig: micelle}).  Note that both the potential and force are continuous at $r=\sigma$.  The crossover in the potential is determined by the radial separation at which the outermost blobs of neighboring micelles fully overlap.  From this we have,
\begin{equation}
\label{eq: sigma}
\sigma = \frac{2 R_0}{1+\sqrt{\pi/p}} \ ,
\end{equation}
and 
\begin{equation}
\xi_{max}=\sqrt{\frac{\pi}{p}} \sigma \ . 
\end{equation}
This pair potential has been shown to agree well with scattering experiments in dilute solutions of copolymer micelles \cite{stellbrink_jphys_04, stellbrink_prl_05} and star polymers \cite{likos_lowen_prl_98}.

To compute the free-energy cost of coronal overlap for micelles in the various structures, we perform lattice sums of the potential in eq. (\ref{eq: Ur}) numerically.  Because $U(r)$ falls off exponentially with distance, the sums are truncated to include a finite number of lattice neighbors, yielding a numerical precision of better than one part in $10^{8}$. Unlike the case of pure Yukawa systems which are characterized by a single length scale, the inter-micelle potential depends explicitly on two separate length scales, namely, $\sigma$ and $\xi_{max}$.  To analyze the influence of the inter-micelle potential on equilibrium structure, in Figure \ref{fig: Upref} we plot the preference of $U(r)$ for either the bcc or fcc structures as a function of $p$ and the {\it reduced density}, $\eta \equiv (\pi/6) n \sigma^3$.  At low densities, $\eta \lesssim 0.7$, micelles in either lattice interact via the long-distance portion of the potential.  In this regime, lattice geometry dictates that the potential energy of fcc structure is exponentially smaller than that of the bcc structure due to its more distant nearest-neighbors.  From studies of Yukawa systems \cite{hone_jcp_83, kremer_prl_86, robbins_jcp_88}, it is known that the bcc structure is generically favored at high densities, and this fact accounts for the $p$-dependent crossover in preferred structure shown in Figure \ref{fig: Upref} for low aggregation number ($p \lesssim 100$).  Near to $\eta \approx 0.7$ nearest-neighboring micelles reach the short-distance, logarithmic regions of $U(r)$, and the crossover density becomes only weakly dependent on $p$.  The crossover from preferred fcc to bcc structures with increased density is a rather generic property of repulsive systems, qualitatively insensitive to the precise form of the inter-micelle potential.

We compute the lattice entropy of competing structures by considering the harmonic free-energy cost of displacing micelles by $\Delta \rv^{\alpha}$ from their equilibrium positions, $\rv_{\alpha}$,
\begin{equation}
{\cal H}_{fluct}=\frac{1}{2} \sum_{\alpha \beta; ij}  \Delta r_i^{\alpha} D_{ij}^{\alpha \beta}  \Delta r_j^{\beta} \ ,
\end{equation}
where the $i$ and $j$ label Cartesian directions, $\alpha$ and $\beta$ label lattice sites and $D_{ij}^{\alpha \beta}$ is the dynamical matrix computed from,
\begin{equation}
\label{eq: D}
D_{ij}^{\alpha \beta}=\frac{\partial^2}{\partial r^{\alpha}_i \partial r^{\beta}_j} \sum_{\gamma \neq \delta} U(|\rv_{\gamma}-\rv_{\delta}|) \ ,
\end{equation}
with $\partial/\partial  r^{\alpha}_i $ the derivative with respect to the $i$ component of the position of the $\alpha$th lattice site.  This matrix can be be diagonalized in Fourier space to find the eigenvalues, $\lambda_i (\kv)$, associated with the three phonon polarizations at wavevector $\kv$.  In the harmonic approximation \cite{born_huang} the equilibrium mean-square displacement can be computed by
\begin{equation}
\label{eq: dr}
\langle |\Delta \rv|^2 \rangle = \frac{k_B T}{N_m} \sum_{i=1}^3 \sum_{\kv \in \{ \Gv \}} \frac{1}{\lambda_i(\kv)} \ ,
\end{equation}
where $\{ \Gv \}$ is the set of reciprocal lattice vectors corresponding to a lattice of $N_m=nV$ micelles.  The free-energy contribution (per micelle) from positional fluctuations in the harmonic approximation is computed by,
\begin{equation}
\label{eq: Ffl}
\frac{F_{fluct}}{k_B T} = \frac{1}{2 N_m} \sum_{i=1}^3 \sum_{\kv \in \{ \Gv \} } \ln \Big(\frac{\lambda_i(\kv)}{k_B T} \Big) \ . 
\end{equation} 

Although the potential in eq. (\ref{eq: Ur}) and its first derivative are continuous over the full range, the second derivative is discontinuous at $r=\sigma$, which would enter the computed phonon spectrum through eq. (\ref{eq: D}).  Hence, the form of $U(r)$ proposed in ref. \cite{likos_lowen_prl_98}, would lead to certain anomalies in the dependence of the fluctuation contribution to the free energy on micelle density, $n$.  To circumvent this issue while maintaining some of the detailed dependence of the vibrational spectrum on the inter-micelle potential, we use only the long-distance, Yukawa form of $U(r)$ -- even down to $r \leq\sigma$ -- when computing the dynamical matrix.  We expect that this becomes a poor approximation well above the density at which the nearest-neighbor separation reaches $\sigma$.  Hence, in this model the vibrational properties of the micelle lattices will be identical to those characterized by Robbins, Kremer and Grest for lattice phases of repulsive Yukawa systems \cite{kremer_prl_86, robbins_jcp_88}.

To calculate the dependence $\langle |\Delta \rv|^2 \rangle$ and $F_{fluct}$ on  $p$ and $n$, we compute $D_{ij}^{\alpha \beta}$ using eq. (\ref{eq: D}) and the long-distance form of $U(r)$, numerically performing lattice sums.  The eigenvalues $\lambda_i(\kv)$ are computed, and then the Fourier sums in eqs. (\ref{eq: dr}) and (\ref{eq: Ffl}) are carried out by summing over a finite set of reciprocal lattice vectors.  More than 300,000 reciprocal lattice points are used to achieve a numerical convergence of $\langle |\Delta \rv|^2 \rangle$ and $F_{fluct}$ to better than one part in $10^{4}$.  

When micelle interactions are sufficiently weak, lattice fluctuations will lead to a melting of the ordered phases.  To assess this instability to the fluid micelle phase, we apply the Lindemann criterion for crystal melting, which defines the melting of lattice order to occur when $\langle |\Delta \rv|^2 \rangle^{1/2} /r_{nn} \geq \alpha_m$, where $R_{nn}$ is the nearest-neighbor lattice spacing and $\alpha_m$ is a numerical constant less than 1 \cite{lindemann}.  Here, we use a Lindemann parameter of $\alpha = 0.19$, which was found to yield good agreement with the Hansen-Verlet criterion for melting \cite{hansen_verlet} in numerical studies of repulsive Yukawa systems \cite{robbins_jcp_88}.  As was found in these studies, the soft-phonon modes of the bcc crystal in the harmonic model are completely unstable to thermal fluctuations for sufficiently low densities (i.e. $n \xi_{max}^3 \lesssim 0.002$), and therefore, the  model discussed above cannot be used to compute the vibrational properties of the bcc structures at these densities.  In this regime, we use an extrapolation of $F_{fluct}$ from the stable bcc phase at higher densities to model the entropic contributions to the free energy of the lower density {\it fluid} phase.

\section{Model Results and Micelle Solution Thermodynamics}

\label{sec: results}

\begin{figure}
\resizebox{3.25in}{!}{\includegraphics{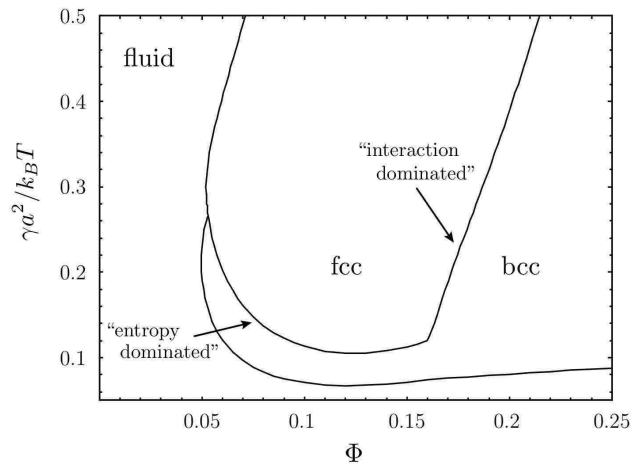}}
\caption{Equilibrium phase behavior for solution of symmetric copolymers $N_c=N_b=50$.  The regions of stable fcc and bcc phases are indicated, as well as the region where the crystal structures are unstable to the micelle fluid phase as determined by the Lindemann criterion (see text).  All phase transitions are first order.  Portions of the fcc-bcc phase boundary are labeled according to the physical mechanism which dominates the thermodynamics as discussed in the text.}
\label{fig: phase_50_50}
\end{figure}

In this section we discuss the mean-field results of the model for copolymer micelle solutions introduced in Sec. \ref{sec: model}.  Calculations are performed for fixed polymer content, $\Phi$, and polymer composition, $N_c$ and $N_b$.  Temperature dependence of the solvent selectivity is encoded in the core-corona surface tension, $\gamma$.  For simplicity we assume that $a=\rho_0^{-1/3}$.  The total free energy (per micelle), $F_{tot}$ is the sum of internal micelle free-energy contributions -- eqs. (\ref{eq: Fc}), (\ref{eq: Fint}) and (\ref{eq: Fb}) -- along with the inter-micelle free-energy contributions, $U$ and $F_{fluct}$.  For fixed solution conditions, the intensive free energy per chain, $F_{tot}/p$, is minimized over micelle aggregation number, $p$.  Mean-field free energy calculations for ordered fcc and bcc micelle structures are compared to construct phase diagrams.  We begin with a detailed discussion of micelle thermodynamics for solutions of symmetric copolymer chains.

\subsection{Symmetric Chains}

\begin{figure}[t]
\resizebox{3.25in}{!}{\includegraphics{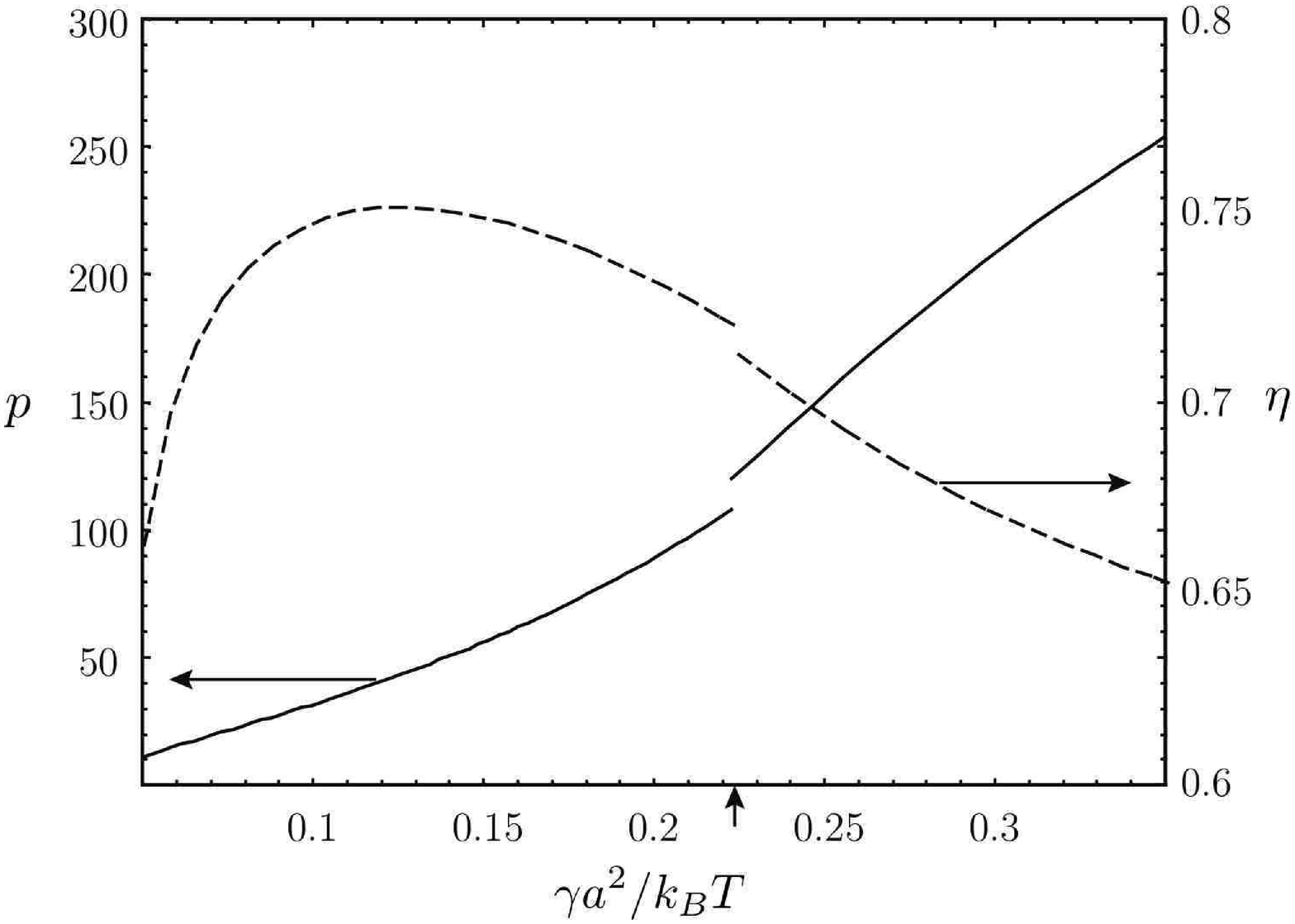}}
\caption{Mean-field values of aggregation number, $p$, (solid line) and reduced density, $\eta=(\pi/6) n \sigma^3$ for solutions of symmetric chains ($N_c=N_b=50$) of polymer fraction $\Phi=0.175$.  The arrow at $\gamma a^2/k_B T = 0.222$ indicates the location of the bcc-fcc phase boundary.  Hence, for values of surface tension smaller (larger) than this the plotted quantities correspond to the equilibrium bcc (fcc) structure. }
\label{fig: p_e}
\end{figure}

In Figure \ref{fig: phase_50_50} we show the predicted phase behavior of copolymer solutions for symmetric molecules with $N_c=N_b=50$.  At sufficiently large core-corona surface tension, $\gamma a^2 /k_B T \gtrsim 0.08$, and polymer volume fraction, $\Phi \gtrsim 0.05$, the micelle solution orders into either the fcc or bcc structures.  The former structure is preferred at lower $\Phi$, while the stable bcc structure is predicted at higher polymer fractions and at lower $\gamma$, before crystal melting.  The topology of the mean-field phase diagram for the ordered micelle phases is consistent with experimental observations of symmetric diblock solutions, excepting here that the temperature dependence of solvent quality is encoded in $\gamma$ so that direct comparison to experiment is not possible at present.  Nonetheless, the predicted fcc$\rightarrow$bcc$\rightarrow$fluid micelle phase sequence with increasing temperature is a generic feature observed for a variety of neutral copolymer solutions \cite{lodge_macro_02,  bang_lodge_jcp_04, bang_lodge_prl_04_1}.  Moreover, we note that many systems (though not all) exhibit the fcc-bcc structural transition with increasing concentration as predicted by the mean-field model (see. e.g. ref. \cite{bang_lodge_jcp_04}).

To put the thermodynamics of lattice choice in the context of equilibrium micelle structure we plot the dependence of mean aggregation and reduced density as functions of the core-corona surface tension in Figure \ref{fig: p_e}.  Since $\gamma$ is assumed to be a decreasing function of $T$, this phase-space trajectory describes a thermotropic fcc-bcc transition.  Here, we see that $p$ has a roughly linear dependence on $\gamma$, not unlike the case where inter-micelle effects are ignored \cite{birshtein_zhulina_poly_89}.  In the range $\gamma a^2/k_B T > 0.12$, $\eta$ is a monotonically decreasing function of $\gamma$.  This stems from the fact that micelle density, $n$, is inversely proportional to $p$, while at large aggregation numbers $\sigma^3 \sim R_0^3$ grows slower than linearly with increased $p$ (for example, $R_0^3 \sim p^{3/5}$ for hairy micelles).  Hence, we have the generic property that at large aggregation numbers increasing core-corona repulsion makes the system less dense on the scale of the intermicelle repulsion, $\sigma$.  Indeed, the rise in density with increased temperature is observed in scattering experiments from micelle lattices in solution \cite{bang_lodge_prl_04_1}.  From eq. (\ref{eq: sigma}) we have $\sigma \sim R_0 p^{1/2}$ for small $p$, thus $\eta$ decreases to 0 as $p\rightarrow 0$, which accounts for the decrease of the reduced micelle density as $\gamma a^2 /k_B T$ is decreased below $0.12$.  For a broad range of conditions in this model, when $p$ is reduced below roughly 20 chains per micelle, ordered structures are unstable to the fluid phase, in agreement with the order of magnitude predictions for the critical chain number for ordering in solutions of star polymers \cite{witten_pincus_cates_epl_86}.

\begin{figure}
\resizebox{3.0in}{!}{\includegraphics{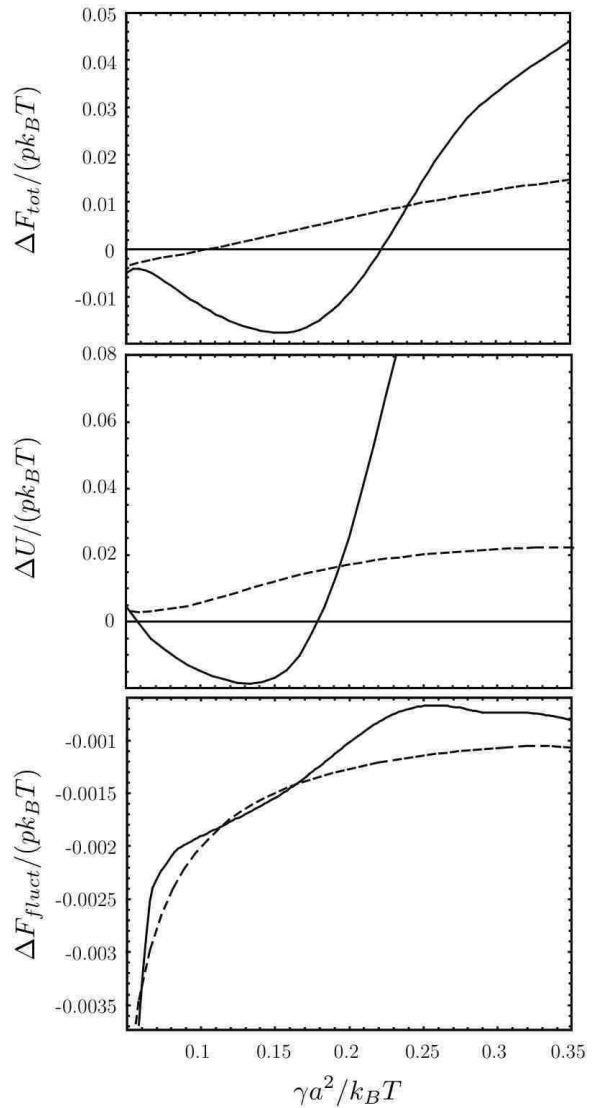}}
\caption{Differences between bcc and fcc the total, $\Delta F_{tot}$, interaction $\Delta U$, and fluctuation, $\Delta F_{fluct}$ free energies of solutions of symmetric chains ($N_c=N_b=50$) as a functions of core-corona surface tension.  The solid line corresponds to results at $\Phi=0.175$ while the dashed line corresponds to $\Phi=0.125$.}
\label{fig: deltaF}
\end{figure}

In principle, both the results that the system effectively becomes more dense and the number of chains per micelle decreases  with {\it decreasing} core-corona surface tension point to the bcc structure as the stable micelle lattice at high temperatures.  The former effect drives an inter-micelle potential preference for bcc (see Fig. \ref{fig: Upref}) while the latter effect suggests that lattice fluctuations stabilize this phase as micelles dissolve, since the strength of effective micelle repulsions is proportional to $p^{3/2}$.  In order to discriminate between these two distinct physical mechanisms for the thermotropic fcc-to-bcc transition -- driven by either lattice potential or lattice entropy -- we plot the difference between $F_{tot}$, $U$ and $F_{fluct}$ (per polymer chain) for the bcc and fcc structures as a function of $\gamma$ in Figure \ref{fig: deltaF}.  As is the case for pure Yukawa systems \cite{robbins_jcp_88}, the relatively large entropy of the fluctuations in the bcc structure always favors that structure over fcc, so that $\Delta F_{fluct} < 0$ for all conditions.  However, this entropic preference amounts to roughly $10^{-3} k_B T$ per chain, which should be compared to differences in lattice potential, which tend to vary on the scale of $10^{-2} k_B T$ per chain.

Consider first the lower polymer fraction case, where $\Phi=0.125$.  Over the range of $\gamma$ plotted in Figure \ref{fig: deltaF}, $\Delta U$ is always positive indicating that interactions favor the fcc structure for this polymer fraction.    Below $\gamma a^2/k_B T = 0.105$, however, the bcc structure is thermodynamically stable over the close-packed lattice until melting for $\gamma a^2/k_B T \le 0.067$.  In this regime of weak core-corona repulsion, mean-aggregation of micelles is sufficiently small ($p \lesssim 20$) that differences in $F_{fluct}$ which favor the bcc lattice become comparable to the preference of micelle interactions favoring the fcc structure.  Hence, we view this thermotropic fcc-to-bcc transition to be of entropic origin, resulting from the abundance of soft-phonon modes in the bcc crystal.

\begin{figure}
\resizebox{3.0in}{!}{\includegraphics{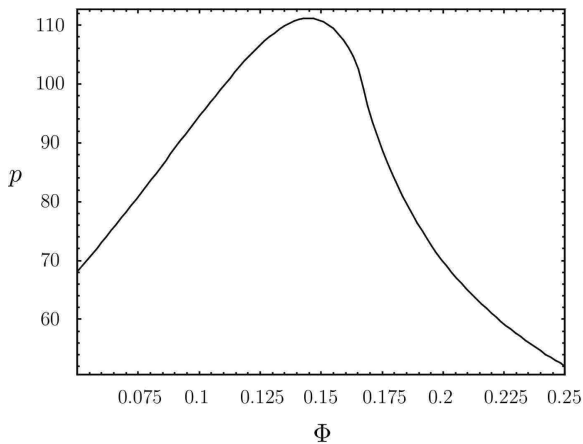}}
\caption{Dependence mean aggregation on $\Phi$ for symmetric copolymer solutions at fixed $\gamma a^2/k_B T$ and fixed bcc structure.}
\label{fig: p_phi}
\end{figure}

\begin{figure*}
\resizebox{6.5in}{!}{\includegraphics{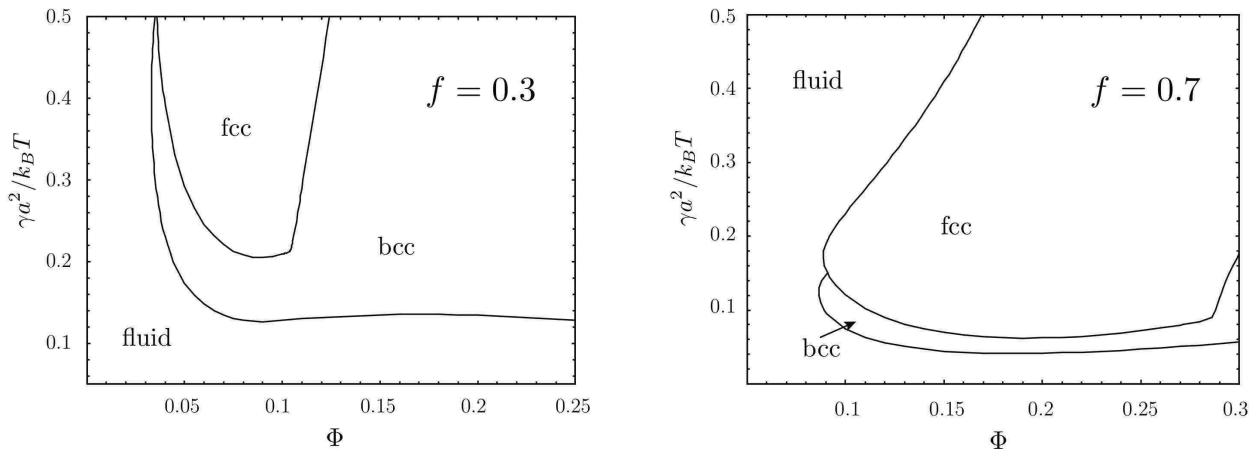}}
\caption{Mean-field phase diagrams for asymmetric $N=100$ copolymer micelle solutions.  The phase are labeled as in Figure \ref{fig: phase_50_50}.  Qualitatively, the $f=0.3$ corresponds to a typical phase diagram for ``hairy" micelles, while the $f=0.7$ represents the ``crew-cut" micelle case.}
\label{fig: phase_hairy_crew}
\end{figure*}

We now turn to the thermotropic fcc-to-bcc transition at $\Phi=0.175$ shown also in Figure \ref{fig: deltaF}.  In contrast to the lower concentration case we find a high temperature region, $0.057 \le \gamma a^2/k_B T \le 0.179$, wherein micelle interactions prefer the bcc structure over fcc.  Indeed, from Figure \ref{fig: p_e}, we find that the system becomes more dense as $\gamma$ is lowered, so that $\eta \gtrsim 0.7$ in this region.  While the generic preference of $F_{fluct}$ for the bcc structure may further stabilize this lattice over the fcc structure, the plot of $\Delta U$ for this concentration demonstrates that the differences in micelle interactions dominate the entropic contribution to the free energy over much of the range of the stable bcc lattice (note that the bcc lattice melts below $ \gamma a^2/k_B T = 0.077$ for this polymer fraction).  Therefore, in this higher concentration regime the structural transition is more appropriately attributed to micelle interactions which favor the bcc structure as the system becomes effectively more dense.

Micelle aggregation is a decreasing function of temperature, which in turn weakens micelle interactions and increases the number density of micelles.  In Figure \ref{fig: phase_50_50}, crossover from entropy-driven to interaction-driven fcc-to-bcc structural transitions is marked by a cusp in phase boundary at $\Phi=0.160$ and $\gamma a^2/k_B T=0.120$.  For micelles with $p \gtrsim 100$, there is effectively a critical value of reduced density, $\eta^* \simeq 0.7$, above which the long range of inter-micelle repulsion leads to a stabilization of the bcc structure.  Since $n$ is linearly proportional to the polymer fraction, increasing $\Phi$ allows the system to pass into the bcc-preferred region, $\eta \gtrsim 0.7$ at larger aggregation number, or alternatively, larger $\gamma$.  This accounts for the $\Phi$-dependence of the fcc-bcc boundary.  For $\Phi \le 0.160$, the weakening of the micelle potential with the loss of chain aggregation supersedes the density-driven transition, leading to the entropic stabilization of the bcc phase near to the melting point.  Therefore, we find that the interaction-driven fcc-to-bcc transition tends to occur at higher $\Phi$, with the entropically-driven structural transition occurring at lower polymer concentrations.

Having analyzed the thermotropic transition in detail, we now briefly consider the case of fixed $\gamma$.  Clearly, the discussion above implies the low-$\Phi$ stability of the fcc structure and the subsequent stability of bcc at higher concentrations due to the micelle interactions.  However, the predicted fluid$\rightarrow$bcc$\rightarrow$fcc$\rightarrow$bcc phase sequence with increasing $\Phi$ predicted for the range $0.105< \gamma a^2/k_B T<0.265$, implies an unusual thermodynamic dependence of micelle aggregation on $\Phi$ (shown in Figure \ref{fig: p_phi}).  While at low concentrations $p$ will be insensitive to $\Phi$, when micelles begin to interact strongly the system adjusts the micelle structure in order to reduce the free energy cost of micelle overlap.  In the concentration regime where micelles interact via the Yukawa portion of $U(r)$, the mean aggregation increases roughly linearly with $\Phi$ in order to maintain a density where nearest-neighbor micelles do not overlap strongly.  Once the system reaches a density where nearest neighbors begin to interact via the short-distance, logarithmic portion of $U(r)$, this trend reverses and aggregation then {\it decreases} with increased $\Phi$.  This effect not only increases the {\it density} of the lattice but also reduces the {\it strength} of repulsions.  The non-monotonic dependence of $p$ on $\Phi$ shown in Fig. \ref{fig: p_phi} explains the re-entrant bcc phase at high concentration, since this structure is generically preferred at high density and low $p$.  This unusual result that $p$ first increases with $\Phi$ and then decreases, is a fundamental consequence of the fact that $U(r)$ is finite-ranged and ``ultra-soft" meaning that it should remain finite down to very small separations.  To prevent significant overlap, the system first slightly increases $p$, reducing density.  Once strong overlap becomes unavoidable, it becomes preferable to decrease $p$, since the free-energy cost of fusing two micelles cannot be significantly larger than creating a single micelle of $2p$ chains.

\subsection{Copolymer Composition:  ``Hairy" vs. ``Crew-Cut" Micelles}

We now consider the influence of copolymer composition on the predictions of the mean-field model.  Defining $f$ as the fraction of chain which is insoluble,
\begin{equation}
f=\frac{N_c}{N_c+N_b} \ ,
\end{equation}
we show the mean-field phase behavior for micelle solutions of corona-rich, $f=0.3$, and core-rich, $f=0.7$, copolymers in Figure \ref{fig: phase_hairy_crew}.  Qualitatively, these two cases correspond to the ``hairy" and ``crew-cut" descriptions of micelle structure, respectively.  More quantitatively, we can characterize the micelle structure by the ratio of the height, $H_b=R_0-R_c$, of the coronal block to the core size.  At $\Phi=0.2$ and $\gamma a^2/k_B T = 0.2$, we have $H_b/R_c= 2.734$ for $f=0.3$ and $H_b/R_c= 0.768$ for $f=0.7$.  Another significant difference is the mean aggregation since $p$ is an increasing function of $N_c$:  $p=28.5$ and $p=188$, for the ``hairy" and ``crew-cut" cases, respectively.  These theoretical phases diagrams recapitulate the observations of Lodge and coworkers that structural order of ``hairy" (``crewcut") micelles is principally of bcc (fcc) type \cite{lodge_macro_02}.

It is the latter distinction in micelle aggregation for these two cases which accounts for the broad differences in the mean-field phase diagrams in Figure \ref{fig: phase_hairy_crew}.  Scaling analysis of $F_0$ suggests a roughly linear dependence of mean-aggregration number on the length $N_c$, which justifies the large differences in $p$ for the two cases.  Since a {\it smaller} micelle aggregation implies a {\it larger} micelle density, for given solution conditions ``hairy" micelles are characterized by highly overlapping coronal brushes while the ``crew-cut" micelles tend to form less dense lattices.  In the low temperature regimes for which ($p\gtrsim 50$ this implies that micelle interactions tend to stabilize the bcc lattice in the former case and fcc in the latter.  Clearly, the effect of micelle structure does not drastically alter the entropic stabilization of the bcc phase near to lattice melting.

\begin{figure}
\resizebox{3.in}{!}{\includegraphics{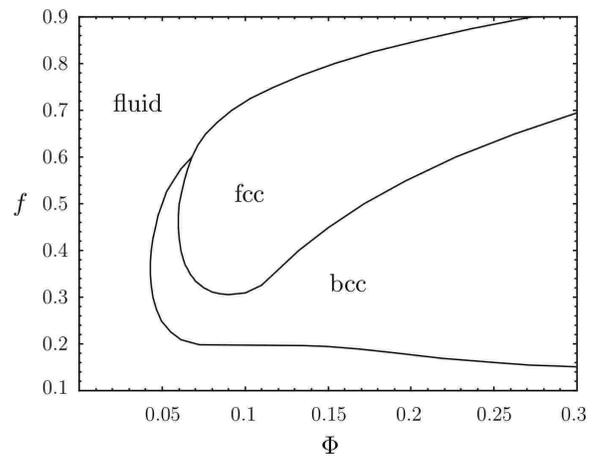}}
\caption{Mean-field phase behavior of $N=100$ copolymer solutions as a function of chemical composition, $f$, and polymer content $\Phi$, for fixed solvent-core repulsion, $\gamma a^2/k_B T=0.2$.}
\label{fig: phase_f}
\end{figure}

In Figure \ref{fig: phase_f} we plot the fixed $\gamma a^2/k_B T=0.2$ phase behavior of $N=100$ copolymer solutions as a function of $\Phi$ and $f$.  Here again, the core-rich copolymer micelles favor the close-packed structure, while the corona-rich micelles favor bcc.  This phase diagram reveals that the re-entrant bcc $\rightarrow$ fcc $\rightarrow$ bcc lyotropic phase sequence is a feature of copolymer solutions over a broad range of chemical compositions.  Although the ``hairy" vs. ``crew-cut" distinction broadly describes ordered-phase thermodynamics of micelle solutions, we find that equilibrium structure cannot be strictly determined by $f$ (or even $H_b/R_c$).  In fact, basic considerations of micelle interactions dictate that ordered-phase thermodynamics of micelles must be {\it insensitive} to the size of the core domain, $R_c$.  Micelle interactions govern both repulsive and entropic free-energy costs of lattice formation, and from eq. (\ref{eq: Ur}) we see that $U(r)$ depends only the number of chains per micelle and the effective micelle size, $\sigma$.  Certainly, this must be the case, given that the self-similar statistics governing the monomer density in the coronal brush domain are independent of the length of coronal chains, $N_b$ \cite{daoud_cotton_jphys_82}.  Therefore, since neighboring micelles interact through overlap of coronal domains where the monomer distribution is governed purely by $p$ and $r$, such interactions cannot encode information about the relative sizes of the core domain.  Or put another way, at the tips of their coronal brushes micelle interactions cannot ``know" how far away the core domain is from the outermost edge of the micelle.  According to a mean-field description of micelle solutions, differences in ordered-phase thermodynamics between the ``hairy" and ``crew-cut" cases are more properly attributed to differences in solvation properties of the copolymers of differing compositions, rather than the relative size of core and coronal domains.

\section{Discussion}

\label{sec: discussion}

In the previous sections we have presented and analyzed a mean-field model for the thermodynamics of concentrated solutions of neutral block copolymers.   This analysis unifies a broad range of observed phenomena regarding structural transitions in lattice phases of micelles.  In particular, the mean-field analysis elucidates the physical mechanisms responsible for phase transitions between ordered states of micelles.  In contrast to other classical particle systems, such as charged-stabilized colloids \cite{hone_jcp_83} or even hard spheres \cite{alder_hoover_young_jcp_68}, modeling the thermodynamics of micellar systems has the additional complication that the internal structure of the micelles themselves establishes equilibrium with the inter-micellar degrees of freedom.  

This fact underlies the basic phenomenology of concentrated solutions of copolymer micelles.  As temperature is raised and solvent selectivity decreases, the mean value of $p$ decreases.  Since the strength of inter-micelle repulsions grows as $p^{3/2}$, the decrease in micelle aggregation leads to a rise in the relative importance of the entropic contributions to lattice free energy resulting from phonon fluctuations.  As the generic arguments of Alexander and McTague dictate \cite{alexander_mctague_prl_78}, this effect promotes the stability of the bcc structure near to the micelle lattice melting.  However, we find that even far from the regimes where entropic considerations are important -- when solvent-selectivity is strong and $p$ is large -- micelle solutions exhibit broad ranges of bcc stability.  In these regions of phase space, stability of the bcc structure over the close-packed fcc structure can be attributed to micellar interactions which are of sufficiently long length scale.

While the present model effectively describes the thermodynamics associated with the fcc-bcc structural transitions in micelle solutions, the approach is limited in regard to other equilibrium structures.  In particular, the star-polymer potential of eq. (\ref{eq: Ur}) has been shown to stabilize non-cubic and even diamond lattice structures \cite{likos_lowen_prl_99, likos_jphys_02}.  Furthermore, studies of similar soft-molecular systems suggest the stability of the relatively open A15 lattice of micelles \cite{ziherl_kamien_prl_00, grason_physrep_06}.  Due to the abundance of thermodynamic studies of copolymer solutions exhibiting the structural fcc-bcc transition, we focus on this particular phenomenon, and consequently, these lattice structures.  Generally, we might expect the stability of other micelle lattice phases at higher polymer concentrations.  More likely, however, is the possibility that other self-assembled morphologies intervene at higher concentration, such as cylindrical, lamellar or network morphologies.  Since the inter-micellar terms in the free energy are necessarily of the same order as the intra-micellar ones, at high concentrations, the system achieves a more efficient packing by a rearrangement of aggregate topology.  Such phase behavior is well-documented in experimental observations of more concentrated copolymer solutions \cite{lodge_macro_02, bang_lodge_jcp_04}.

Other limitations of the model include the fluctuation effects of the micellar structure.  One of these is the swelling of the core domain by solvent at high temperature.  Neutron scattering in dilute solutions of micelles reveals that penetration of solvent into the core domain may reach as high as 30\% by volume near to micelle lattice melting \cite{bang_lodge_prl_04_1}.  While we have not included this effect in our mean-field treatment, we do not expect it to change the qualitative view of ordered phase thermodynamics presented here.  Namely, our principle conclusion that structural fcc-to-bcc transitions are driven by increases in micelle density will remain unchanged.  Indeed, the swelling of the core domain will tend to counteract the decrease in micelle size by loss of polymer chains as temperature is raised.  Thus, the basic prediction that $\eta$ largely {\it increases} as temperature rises will hold in the case that solvent is permitted to swell the micelle core.
 
A second type of fluctuation effect absent from the present model are equilibrium fluctuations in micelle aggregation number, $p$, leading to a certain amount of polydispersity in micelle size.  As in micellar phases of copolymer/homopolymer blends \cite{leibler_jcp_83}, fluctuations in micelle aggregation above the critical micelle concentration can be estimated to be on the order of 10\% or less.  While the degree of polydispersity in the size of micelles will be even smaller than this, it is not clear how even this modest level of polydispersity should affect the relative stability of various ordered phases.  Certainly, a broad distribution in micelle size and aggregation number will frustrate periodic order, but given that copolymer solutions are known to form well-ordered structures over a broad range of equilibrium conditions we expect this effect may not be so significant for this system.

Finally, we conclude by contrasting the thermodynamics of micellar phases to other spherically-aggregating molecular systems, such as ionic or non-ionic surfactants.  Indeed, surfactant systems exhibit structural transitions between competing ordered structures as concentration and solvent quality are varied \cite{imai_jcp_05}.  However, unlike their polymeric counterparts, aggregates of low-molecular weight surfactants tend be characterized by a fixed size, dictated by molecular-scale packing constraints \cite{israelachvili}.  In contrast, micelles of flexible polymeric amphiphiles adjust their aggregation by a factor 5-10 times over the accessible range of solution conditions.  This molecular flexibility is fundamentally linked to the unusual feature that copolymer micelle lattices become {\it more dense} as temperature is raised and molecular forces are weakened, ultimately driving transitions in the long-ranged structural order of the system.

\begin{acknowledgments}
It is a pleasure to thank J.~Bang, R.~Kamien, P.~Ziherl and C.~Santangelo for helpful discussions.  This
work was supported by NSF Grants DMR01-02459 and DMR04-04507, the Donors of the Petroleum Research Fund, Administered by the American Chemical Society and a gift from L.J. Bernstein.
\end{acknowledgments}

\end{document}